\documentclass[twocolumn,showpacs,preprintnumber
s,amsmath,amssymb,aps, prc]{revtex4}

\begin{document}

\preprint{APS}

\title{\large {$\Delta$ Contribution to $ pp \to 
pp\pi^0$}}
\author{Venkataraya$^{1,2,4}$, Sujith 
Thomas$^{2,5}$ and G. Ramachandran$^{2,3}$}
\affiliation{Vijaya College, Jayanagar, 
Bangalore$^{1}$,\\ G.V.K. Academy, 
Bangalore$^2$,\\ Amrita Vishwa Vidya Peetham, 
Bangalore$^3$,\\ Amrita Vishwa Vidya Peetham, 
Coimbotore$^4$\\ Texas Instruments, Bagmane Tech 
Park, Bangalore$^5$}
\email{venkataraya@gmail.com, 
sugiththomas@gmail.com, gwrvrm@yahoo.com}

\date{\today}
\begin{abstract}
A model independent discussion of 
$\Delta-$production in $pp \to pp \pi^0$ is 
presented and the $\Delta-$contributions to the 
16 $pp \to pp\pi^0$ partial wave amplitudes are 
identified taking the $Ds$ and $Sd$ channels 
also into consideration.
\end{abstract}
\maketitle

\section{INTRODUCTION}

\hspace{0.5cm} The earliest experimental studies 
\cite{1} on meson production in nucleon-nucleon 
collisions were carried out within two decades 
of Yukawa's theoretical
prediction \cite{2} of the existence of mesons 
and within only a few years after the 
experimental discovery \cite{3} of the charged 
and the neutral pions.
Phenomenological models were used \cite{4} to 
interpret the data and notations like 
$\sigma_{11},\sigma_{01},\sigma_{01}(np)$ and 
$\sigma_{10}(d)$
were used for total cross-section, where the 
indices referred to the initial and final 
isospin states of the two nucleons.
The cross-sections $\sigma(pp \to pp\pi^0)$ and 
$\sigma(pp \to d\pi^+)$ were denoted by 
$\sigma_{11}$ and $\sigma_{10}(d),$ while 
$\sigma_{10}(np)$
referred to $\sigma(pp \to np\pi^+) -\sigma_{11} 
$ and $\sigma_{01} = \sigma(np \to pp\pi^-) + 
\sigma_{11}(np \to nn\pi^+) - \sigma_{11}.$
Polarization measurements have also been 
suggested quiet early \cite{5} and it was noted 
\cite{6} that the ratio of $\sigma(pp \to 
np\pi^+)$ to
$\sigma_{10}(d),$ appeared to be more than twice 
as large as predicted.
The pion was assumed to be produce mainly in the 
$p$ state and the partial waves were classified 
as $Sp, Ss Pp$ and $Ps,$ where the capital 
symbol
refers to the orbital angular momentum $l_f$ 
between the two nucleons in the final state and 
the lower case symbol refers to the orbital 
angular momentum
$l$ of the meson with respect to the final two 
nucleon system.
In the case of $\sigma(pp \to pp\pi^0)$, 
however, it may be noted that the $Sp$ final 
state is ruled out as it implies total angular 
momentum $j=1$
with positive parity, whereas the conservation 
of total angular momentum and parity requires 
the initial two proton state to be $^3S_1$ or 
$^3D_1$
violating the Pauli exclusion principle.

The construction of meson factories with intense 
well-defined proton beams made it possible to 
carry out accurate and kinematically complete 
cross
section measurements and the advent of storage 
rings with electron cooled beams and windowless 
internal gas targets \cite{7} opened up the near 
threshold
region for precise experimental study.
Only a few lowest partial waves are involved at 
threshold, which in turn limit the initial 
partial waves $l_i$, through the conservation of 
total angular
momentum and parity.
Total cross section measurements close to 
threshold for $\sigma(pp \to pp\pi^0)$ \cite{8}, 
$\sigma(pp \to np\pi^+)$ \cite{9} and $\sigma(pp 
\to d\pi^+)$
\cite{10} in the early 1990's led to 
disagreements with theory \cite{11}.
In particular, the total cross section for 
$\sigma(pp \to pp\pi^0)$ was found quite 
surprisingly to be more than $5$ times the then 
existing theoretical
predictions \cite{11}.
This catalyzed theoretical studies \cite{12} in 
model building, while advances in technology 
\cite{13} employing polarized beams and 
polarized targets.
The measurements \cite{14} were compared with 
the predictions of the Julich meson exchange 
model \cite{15} and were also analysed 
empirically using relations
derived by Bilenky and Ryndin \cite{16} for 
total cross sections.
Some experimental measurements \cite{17} 
indicated $Sd$ and $Ds$ contributions, which 
were not considered earlier \cite{14}.
We may also refer to several reviews \cite{18}.

Empirical formulae were derived \cite{19} for 
$NN \to NN\pi$ cross section at the double 
differential level, which on integration led to 
the earlier
result of \cite{16} for total cross section.
It was also shown \cite{20} that it is possible 
to partition empirically the differential cross 
section into the four contributions from the 
initial
singlet $| 0,0 \rangle$ and triplet $| 1,m 
\rangle, m = \pm1,0$  spin states employing the 
technological capabilities at the PINTEX 
facility and
theoretical studies using effective field 
theory.

Kinematically complete measurements of spin 
observables in polarized beam and polarized 
target experiments were reported on neutral 
\cite{21} and
charged \cite{22} pion production.
The Julich model \cite{15} was found to be more 
successful with less complete data \cite{22} on 
$\vec{p}\vec{p} \to d \pi^+$ and
$\vec{p}\vec{p} \to np \pi^+$ than with the more 
complete set of spin observables \cite{21} 
reported in the case of $\vec{p}\vec{p} \to pp 
\pi^0$.
In their reviews \cite{18} Moskal et.al., and 
Hanhart  have both remarked that apart from rare 
cases, it is difficult to extract particular 
piece of
information from the data.

The model independent approach \cite{19,20} 
based on irreducible tensor techniques \cite{23} 
has been employed \cite{24} to analyse the data 
on
$\vec{p}\vec{p} \to pp \pi^0$ , taking into 
consideration all the twelve partial wave 
amplitudes covering the $Ss, Ps$ and $Pp$ 
channels but
neglecting $Sd$ and $Ds$ following Meyer et. al. 
\cite{21}.
The extracted partial wave amplitudes when 
compared \cite{25} with Julich model 
predictions, revealed that\\
i) the $\Delta$ degree of freedom is important 
for quantitative understanding of the reaction 
$pp \to pp \pi^0$ ,\\
ii) the discrepancy between the empirical and 
model estimates was maximum in $^3P_1 \to ^3P_0 
p$ and to a lesser extent in $^3F_3 \to ^3P_2 p$ 
and\\
iii) a need possibly to include higher partial 
waves as well.
One of the short-comings of \cite{25} was 
\cite{26} that the phase ambiguity of the 
threshold $Ss$ amplitude was assumed to be the 
same as that of the
leading $Ps$ amplitudes from initial $^1S_0.$
It was shown in \cite{26}, how this phase 
ambiguity can be overcome by treating the final 
state spin observables in $pp \to \vec{p}\vec{p} 
\pi^0$ with
initially unpolarized protons.

The importance of $\Delta$ contribution have 
also been noted in several earlier studies 
\cite{27}.
In fact,  $NN \to N \Delta $ has a rich spin 
structure \cite{28}.
Of the sixteen amplitudes associated with $NN 
\to N \Delta$, as many as ten are second rank 
tensors \cite{29}.
Augar et.al., \cite{30} have suggested amplitude 
determination from measurements on spin 
observables.
The measurement of analyzing powers has been 
reported \cite{31} with one of the colliding 
protons polarized in neutral pion production.
The charged pion production was considered in 
\cite{32}.

The purpose of the present paper is to focus 
attention on $\Delta$ production in $pp \to pp 
\pi^0$ using the model independent theoretical 
approach
\cite{19,20,28}.

\section{MODEL INDEPENDENT APPROACH TO 
$\Delta-$PRODUCTION}

\subsection{\sc Kinematical Considerations}

\hspace{0.5cm} Let us consider $pp  \to 
p\Delta^+ \to pp \pi^0$ at center of mass energy 
$E,$ where the colliding protons share the 
energy $E$ equally and have momenta ${\bf p}_i$ 
and $-{\bf p}_i$ with ${\bf p}_i = p_i \hat{\bf 
p}_i$ along the $z-$axis.
Let us denote the energies and momenta of 
$\pi^0$ and the two protons in the final state 
as $ (\omega, {\bf q} = q \hat{\bf q});
(E_1, {\bf p}_1 = p_1 \hat{\bf p}_1); (E_2, {\bf 
p}_2 = p_2 \hat{\bf p}_2)$ respectively such 
that

\begin{equation}\label{1}
{\bf p}_1 + {\bf p}_2 + {\bf q} = 0
\end{equation}

We use natural units $ c = \hbar = 1$ and denote 
the masses of the $\Delta$, proton and pion as 
$M_\Delta, M$ and $m$ respectively.
When we envisage $\Delta$ production, the center 
of mass energy gets divided into

\begin{equation}\label{2}
E_\Delta = \frac{E^2 + M^2_\Delta - M^2}{2E}
\end{equation}

\begin{equation}\label{3}
E_p = \frac{E^2 - M^2_\Delta + M^2}{2E}
\end{equation}
of the  and a proton.
Without loss of any generality, we may choose 
events with $E_2 = E_p$ and proceed to discuss 
$\Delta$ contributions to the measured double 
differential cross-section for $pp \to pp\pi^0$.

For the present discussion, we may define an 
invariant mass $W_{\pi N}$ for the $\pi - N $ 
system in the final state as the positive square 
root of
\begin{eqnarray}\label{4}
W_{\pi N}^2 = (E_1 + \omega)^2 - |{\bf p}_1 + 
{\bf q}_1|^2 \\ \nonumber
= (E - E_p)^2 - |{\bf p}_2|^2 \\ \nonumber
= E^2 + M^2 - 2EE_p
\end{eqnarray}
so that $W_{\pi N} = M_\Delta$ corresponds to 
$\Delta $ production.

If ${\bf r}_1 , {\bf r}_2 $ and ${\bf r}_3$ 
denote respectively the instantaneous locations 
of the two protons and the pion in the final 
state, we may define the location of the center 
of mass of the $\pi - N $ system through

\begin{equation}\label{5}
{\bf R}_{\pi N} = \frac{m{\bf r}_3 + M{\bf 
r}_1}{m+M}
\end{equation}
the relative location of the pion with respect 
to nucleon by

\begin{equation}\label{6}
{\bf r} = {\bf r}_3 - {\bf r}_1
\end{equation}
and the relative location of the proton with 
momentum ${\bf p}_2$ with respect to $\pi - N $ 
system by

\begin{equation}\label{7}
{\bf \rho} = {\bf r}_2 - {\bf R}_{\pi N}
\end{equation}
From the above we have

\begin{equation}\label{8}
{\bf r}_3 = {\bf R}_{\pi N} + \frac{M}{m+M} {\bf 
r}
\end{equation}
and

\begin{equation}\label{9}
{\bf r}_1 = {\bf R}_{\pi N} - \frac{m}{m+M} {\bf 
r}
\end{equation}
From \eqref{8} and \eqref{9}, the pion and 
nucleon momenta are given by

\begin{equation}\label{10}
{\bf q} = m \frac{d {\bf r}_3}{dt} = m 
\frac{d{\bf R}_{\pi N} }{dt} + \mu_{\pi N} 
\frac{d{\bf r}}{dt}
\end{equation}

\begin{equation}\label{11}
{\bf p}_1 = M \frac{d {\bf r}_1}{dt} = M 
\frac{d{\bf R}_{\pi N} }{dt} - \mu_{\pi N} 
\frac{d{\bf r}}{dt}
\end{equation}
where $\mu_{\pi N} = \frac{mM}{m+M}$ is the 
reduced mass of the pion nucleon system.
Adding the above two equations gives

\begin{equation}\label{12}
{\bf q} + {\bf p}_1 = -{\bf p}_2 =M \frac{d {\bf 
r}_1}{dt} = (m + M) \frac{d{\bf R}_{\pi N} }{dt}
\end{equation}
Multiplying \eqref{10} by $M$ and \eqref{11} by 
$m$ and subtracting one from the other leads to

\begin{equation}\label{13}
M{\bf q} - m {\bf p}_1 = (m + M)\mu_{\pi N} 
\frac{d{\bf r}}{dt} = (m + M){\bf q^\prime}
\end{equation}
where  

\begin{equation}\label{14}
{\bf q^\prime} = \mu_{\pi N} \frac{d{\bf r}}{dt}
\end{equation}
is the relative momentum between the pion and 
the nucleon given by

\begin{equation}\label{15}
{\bf q^\prime}  = \frac{M{\bf q} - m {\bf 
p}_1}{m + M} = {\bf q} + \frac{m}{m+M} {\bf p}_2
\end{equation}

If $l^\prime$ denotes, the relative orbital 
angular momentum between the $\pi^0$ and the 
nucleon which comes out with momentum ${\bf 
p}_1,$ it is clear
that $l^\prime = 1$ combines with the spin $S_1 
= \frac{1}{2}$ of the nucleon to lead to the 
spin $S_\Delta = \frac{3}{2}$ of the $\Delta.$

\subsection{\sc Reaction Matrix for $pp \to p 
\Delta^+$}

\hspace{0.5cm} The reaction matrix $\mathcal{M}$ 
for $pp \to p \Delta^+$ may be written in the 
form

\begin{equation}\label{16}
\mathcal{M} = \sum_{s_i = 0}^1 \sum_{s=1}^2 
\sum_{\Lambda = |s - s_i|}^{s +s_i} (S^\Lambda 
(s,s_i)\cdot \mathcal{M}^\Lambda (s,s_i))
\end{equation}
where the irreducible spin tensor operators are 
defined following \cite{23} and the 
corresponding tensor amplitudes 
$\mathcal{M}^\Lambda_\mu (s,s_i)$ are 
expressible in terms of the partial wave 
amplitudes $\mathcal{M}^j_{l_2 s; l_i s_i}$, 
which completely take care of the dependence on 
c.m. energy $E.$
We have

\begin{eqnarray}\label{17}
 \mathcal{M}^\Lambda_\mu (s,s_i) = \sum_{l, l_2, 
j} (-1)^{s_i - j} [j]^2 [s]^{-1} W(s_i l_i s 
l_2; j \Lambda)\nonumber \\ \mathcal{M}^j_{l_2 
s; l_i s_i}   (Y_{l_2} (\hat{\bf p}_2) \otimes 
Y_{l_i} (\hat{\bf p}_i) )^\Lambda_\mu 
\hspace{1in} 
\end{eqnarray}
to describe events with $W_{\pi N} = 
\mathcal{M}_{\Delta} $ and $E_2 = E_p$. We use 
the shorthand notation $[j]=\sqrt{2j+1}$. With 
$s_i = 0,1$; $s = 1,2$ and $\Lambda = (s - s_i), 
\ldots (s+s_i)$ one can construct as many as

\begin{equation}\label{18}
N =  \sum_{s_i = 0}^1 \sum_{s=1}^2 \sum_{\Lambda 
= |s - s_i|}^{s +s_i} (2\Lambda + 1) = 32
\end{equation}
irreducible tensor amplitudes.
However, it has been noted \cite{23} that the 
number of non-zero irreducible tensor amplitudes 
get reduced to $16.$
Choosing a right-handed cartesian coordinate 
system with the $z-$axis along ${\bf p}_i$ and 
$y-$axis along ${\bf p}_i \times {\bf p}_2$  
(which may be referred to as Madison frame), it 
follows that $\mathcal{M}^\Lambda_{-\mu} 
(s,s_i)=(-1)^{\Lambda-\mu}\mathcal{M}^\Lambda_{\%
mu} (s,s_i)$ using the property $Y_{l 
-m}(\theta,\phi)=(-1)^{-m}e^{-2im\phi} 
Y_{lm}(\theta,\phi)$. The irreducible tensor 
amplitudes are shown  in Table-I, where the 
number $n$ of the independent amplitudes is also 
shown.

\begin{table}[t]
\caption{A list of non-zero irreducible tensor 
amplitudes for $pp \to p\Delta^+$}.\\
\begin{tabular}{|ccccc|}\hline
$s_i$ & $s$ & $\lambda$ & 
$\mathcal{M}^\Lambda_\mu (s,s_i)$ & $n$\\ \hline
0 & 1 & 1 & $\mathcal{M}^1_{\pm 1}(1,0)$ & 1\\
0 & 2 & 2 & $\mathcal{M}^2_{0}(2,0),\, 
\mathcal{M}^2_{\pm 1}(2,0),\, \mathcal{M}^2_{\pm 
2}(2,0)$ & 3\\
1 & 1 & 0 & $\mathcal{M}^0_{0}(1,1)$ & 1\\
1 & 1 & 1 & $\mathcal{M}^1_{\pm 1}(1,1)$ & 1\\
1 & 1 & 2 & $\mathcal{M}^2_{0}(1,1),\, 
\mathcal{M}^2_{\pm 1}(1,1),\, \mathcal{M}^2_{\pm 
2}(1,1)$ & 3\\
1 & 2 & 1 & $\mathcal{M}^1_{\pm 1}(2,1)$ & 1\\
1 & 2 & 2 & $\mathcal{M}^2_{0}(2,1),\, 
\mathcal{M}^2_{\pm 1}(2,1),\, \mathcal{M}^2_{\pm 
2}(2,1)$ & 3 \\
1 & 2 & 3 & $\mathcal{M}^1_{\pm 
1}(2,1),\,\mathcal{M}^2_{\pm 2}(2,1),\, 
\mathcal{M}^3_{\pm 3}(2,1)$ & 3\\ \hline
 \end{tabular} 
\end{table}

\subsection{\sc Isospin, Spin and Parity 
Considerations}

\hspace{0.5cm} When $\Delta$ is produced, the 
isospin $I_\Delta = \frac{3}{2} $ of the 
$\Delta$ and isospin $I_2 = \frac{1}{2}$ of the 
proton can combine to give the conserved isospin

\begin{equation}\label{19}
I_i = I_f = I = 1
\end{equation}

Likewise, the spin $s_\Delta = \frac{3}{2}$ of 
the $\Delta$ combines with the spin $s_2 = 
\frac{1}{2}$ of the proton ${\bf p}_2$ with 
momentum  in the final state to yield final 
channel spin $s=1,2$ while the initial channel 
spin of the two colliding protons is $ s_i = 
0,1.$
If $l_i$ denotes the relative orbital angular 
momentum between the colliding protons, the 
initial state may be visualized as an isospin 
$I=1$ state,
$| (l_i s_i)jm \rangle$ where $j$ denotes the 
total angular momentum which can take values $ 
|l_i-s_i|, \ldots,(l_i+s_i)$ in steps of $1.$
The Pauli exclusion principle demands that 
$(-1)^{l_i + s_i + I} = -1$ i.e., $(l_i+s_i)$  
must be even.
If $l_2$ denotes the relative orbital angular 
momentum between $\Delta$ and the proton in the 
final state, parity conservation demands that

\begin{equation}\label{20}
(-1)^{l_i} =(-1)^{l_2}
\end{equation}
If we limit ourselves to $l_2 = 0,1$ at 
threshold energies, we have a set of nine 
partial wave amplitudes $\mathcal{M}^j_{l_2 s; 
l_i s_i} = F_\alpha$, where
$\alpha$ collectively denotes

\begin{equation}\label{21}
\alpha = {l_2, s,j, l_i, s_i}
\end{equation}
They are serially numbered, for convenience, as 
$F_1, \ldots F_9$ which are shown in Table-II.
\begin{table}[t]
\caption{Threshold partial wave amplitudes for 
$pp \to p\Delta^+$.}
\begin{tabular}{|cccccc|}\hline
$l_2$ & $s$ & $j$ & $l_i$ & $s_i$ & 
$\mathcal{M}^j_{l_2 s; l_i s_i} = F_\alpha$\\ 
\hline
0 & 2 & 2 & 2 & 0 & $F_1$\\
1 & 1 & 0 & 1 & 1 & $F_2$\\
1 & 1 & 1 & 1 & 1 & $F_3$\\
1 & 1 & 2 & 1 & 1 & $F_4$\\
1 & 1 & 2 & 3 & 1 & $F_5$\\
1 & 2 & 1 & 1 & 1 & $F_6$\\
1 & 2 & 2 & 1 & 1 & $F_7$\\
1 & 2 & 2 & 3 & 1 & $F_8$\\
1 & 2 & 3 & 3 & 1 & $F_9$\\ \hline
\end{tabular}
\end{table}

Each one of the tensor amplitudes 
$\mathcal{M}^\Lambda_\mu (s,s_i)$ may be 
expressed in the form

\begin{eqnarray}\label{22}
\mathcal{M}^2_{0}(2,0) = g_1 = \frac{1}{4\pi} 
F_1 \nonumber \\
\mathcal{M}^0_0(1,1) = g_2 \cos\theta_2 
\nonumber \\
\mathcal{M}^1_{\pm 1}(1,1) = g_3 \sin\theta_2 
\nonumber \\
\mathcal{M}^2_{0}(1,1) = g_4 \cos\theta_2 
\nonumber \\
\mathcal{M}^2_{\pm 1}(1,1) = \pm g_5 
\sin\theta_2 \nonumber \\
\mathcal{M}^1_{\pm 1}(2,1) = g_6 \sin\theta_2 
\nonumber \\
\mathcal{M}^2_{0}(2,1) = g_7 \cos\theta_2 
\nonumber \\
\mathcal{M}^2_{\pm 1}(2,1) =\pm g_8 \sin\theta_2 
\nonumber \\
\mathcal{M}^3_{\pm 1}(2,1) = g_9 \sin\theta_2
\end{eqnarray}
where $g_2, \ldots, g_5$ are invertible linear 
combinations of the four partial wave amplitudes 
$F_2, \ldots, F_5$.
They are given by 

\begin{equation}\label{23}
g_i = \frac{1}{240\pi} \sum_{j=2}^5 
\mathcal{A}_{ij} F_j; i = 2, \ldots, 5
\end{equation}
likewise

\begin{equation}\label{24}
g_i = \frac{1}{240\sqrt{5}\pi} \sum_{j=6}^9 
\mathcal{B}_{ij} F_j; i = 6, \ldots, 9
\end{equation}
where the matrices $\mathcal{A}$ and 
$\mathcal{B}$ are

$$\mathcal{A} = \left(
\begin{array}{cccc}
 20 & 60 & 100 & 0 \\
-10\sqrt{3} & -15\sqrt{3} & 25\sqrt{3} & 0\\
-20\sqrt{2} & 30\sqrt{2} & -10\sqrt{2} & 
12\sqrt{3}\\
10\sqrt{3} & -15\sqrt{3} & 5\sqrt{3} & 6\sqrt{2}
\end{array}
\right);$$
$$\mathcal{B} = \left(
\begin{array}{cccc}
9\sqrt{5} & 45 & 0 & 0\\
-18\sqrt{6} & -6\sqrt{30} & -12\sqrt{5} & 
-24\sqrt{14}\\
27 & 9\sqrt{5} & -2\sqrt{30} & -8\sqrt{21}\\
0 & 0 & 20\sqrt{6} & 2\sqrt{105}
\end{array}
\right)$$

The unpolarized differential cross-section for 
$pp \to p \Delta^+$ may be written as

\begin{equation}\label{25}
\frac{d \sigma_0}{d \Omega_2} = \frac{1}{4} 
Tr(\mathcal{M}\mathcal{M}^\dag)
\end{equation}

The differential cross-section given by 
\eqref{25} may explicitly be written in terms of 
the nine $pp \to p \Delta^+$ amplitudes as

\begin{equation}\label{26}
\frac{d \sigma_0}{d \Omega_2} = \frac{1}{4} 
\sum_{\Lambda, \mu, s, s_i} 
|\mathcal{M}^\Lambda_\mu (s,s_i)|^2 = a + 
b\cos^2 \theta_2
\end{equation}
where

\begin{equation}\label{27}
a = \frac{1}{4}(|g_1|^2 + 2|g_3|^2 + 2|g_5|^2 + 
2|g_6|^2 + 2|g_8|^2 + 2|g_9|^2)
\end{equation}

\begin{equation}\label{28}
b = \frac{1}{4}(|g_2|^2 - 2|g_3|^2 + |g_4|^2 - 
2|g_5|^2 - 2|g_6|^2 + |g_7|^2 - 2|g_8|^2 - 
2|g_9|^2)
\end{equation}

\subsection{\sc Final state}

\hspace{0.5cm} The three-particle final state of 
$pp\pi^0$ results when $\Delta^+$ decays into a 
proton with momentum ${\bf p}_1$ and pion with
momentum ${\bf q}$ such that ${\bf q^\prime}$ 
represents the relative momentum between the 
pion and the proton given by \eqref{14}.
The relative momentum of the proton produced 
along with  $\Delta^+$ with respect to center of 
mass $\pi -N$ system is readily seen
to be

\begin{equation}\label{29}
\frac{(M+m)m}{2M+m} \frac{d({\bf r}_2 - {\bf 
R}_{\pi N})}{dt} = {\bf p}_2
\end{equation}

The double differential cross-section is then 
given by 

\begin{equation}\label{30}
\frac{d^2 \sigma_0}{d^3 q^\prime d \Omega_2} = 
\frac{q^\prime}{W_{\pi N}} \frac{d^2 
\sigma_0}{dW_{\pi N} d \Omega_2}
\end{equation}

\section{Contribution to  $ pp \to pp\pi^0$ 
amplitudes}

\hspace{0.5cm} To identify the connection 
between the  $ pp \to p\Delta^+$ amplitudes 
employed above and  $ pp \to pp\pi^0$  
amplitudes used earlier \cite{26}, we observe 
that the reaction matrix $\mathcal{M}$ for  $ pp 
\to pp\pi^0$ was written earlier in the form

\begin{equation}\label{31}
\mathcal{M} = \sum_{s_i, s_f = 0}^1 
\sum_{\lambda = |s_i - s_f|}^{s_f + s_i} 
(S^\lambda (s_f , s_i) \cdot \mathcal{M}^\lambda 
(s_f , s_i))
\end{equation}
in terms of the irreducible tensor operators 
$S^\lambda_\mu (s_f , s_i)$  of rank $\lambda$ 
expressing the transition from the initial spin 
state $s_i$ to the final spin state $s_f$ and 
the irreducible tensor amplitudes 
$\mathcal{M}^\lambda_\mu (s_f , s_i)$.
The irreducible tensor amplitudes are 
expressible in terms of the partial wave 
amplitudes $\mathcal{M}^j_{l(l_f s_f)j_f ;l_i 
s_i} (E,W)$ as shown in Table-I of \cite{9}, 
where the $S_s, P_s$ and $P_p$ channels were 
considered.
Including $D_s$ and $S_d$ channels, the sixteen 
partial wave amplitudes are presented here in 
Table-III.\\

\begin{table}[t]
\caption{ List of partial wave amplitudes for 
the reaction $ pp \to pp\pi^0.$}
\begin{tabular}{|cccc|}\hline
Initial $pp$ & Type & Final $pp\pi^0$ & Partial 
wave amplitudes \\
state & & state & $\mathcal{M}^j_{l(l_f s_f)j_f 
;l_i s_i}$ \\ \hline
$^3 P_0$ & $S_s$ & $^1 S_0, s$ & 
$\mathcal{M}^0_{0(00)0;11} = f_1$ \\
$^1 S_0$ & $P_s$ & $^3 P_0, s$ & 
$\mathcal{M}^0_{0(11)0;00} = f_2$ \\
$^1 D_2$ &  & $^3 P_2, s$ & 
$\mathcal{M}^2_{0(11)2;20} = f_3$ \\
$^3 P_0$ & $P_p$ & $^3 P_1, p$ & 
$\mathcal{M}^0_{1(11)1;11} = f_4$ \\
$^3 P_2$ &  & $^3 P_3, p$ & 
$\mathcal{M}^2_{1(11)1;11} = f_5$ \\
$^3 P_2$ &  & $^3 P_2, p$ & 
$\mathcal{M}^2_{1(11)2;11} = f_6$ \\
$^3 F_2$ &  & $^3 P_1, p$ & 
$\mathcal{M}^2_{1(11)1;31} = f_7$ \\
$^3 F_2$ &  & $^3 P_2, p$ & 
$\mathcal{M}^2_{1(11)2;31} = f_8$ \\
$^3 P_1$ &  & $^3 P_0, p$ & 
$\mathcal{M}^1_{1(11)0;11} = f_9$ \\
$^3 P_1$ &  & $^3 P_1, p$ & 
$\mathcal{M}^1_{1(11)1;11} = f_{10}$ \\
$^3 P_1$ &  & $^3 P_2, p$ & 
$\mathcal{M}^1_{1(11)2;11} = f_{11}$ \\
$^3 F_3$ &  & $^3 P_2, p$ & 
$\mathcal{M}^3_{1(11)2;31} = f_{12}$ \\
$^3 P_2$ & $D_s$ & $^1 D_2, s$ & 
$\mathcal{M}^2_{0(20)2;11} = f_{13}$ \\
$^3 F_2$ &  & $^1 D_2, s$ & 
$\mathcal{M}^2_{0(20)0;31} = f_{14}$ \\
$^3 P_2$ & $S_d$ & $^1 S_0, d$ & 
$\mathcal{M}^2_{2(00)0;11} = f_{15}$ \\
$^3 F_2$ &  & $^1 S_0, d$ & 
$\mathcal{M}^2_{2(00)0;31} = f_{16}$ \\ \hline
\end{tabular}
\end{table}
In this scheme, the centre of mass of the two 
protons is given by 

\begin{equation}\label{32}
{\bf R}_{12} = \frac{M {\bf r}_1 + M {\bf 
r}_2}{2M} = \frac{1}{2} ({\bf r}_1 + {\bf r}_2)
\end{equation}
The above together with

\begin{equation}\label{33}
{\bf r}_{12} = {\bf r}_1 - {\bf r}_2
\end{equation}
leads to

\begin{equation}\label{34}
{\bf r}_{1} = {\bf R}_{12} + \frac{1}{2} {\bf 
r}_{12}
\end{equation}
and

\begin{equation}\label{35}
{\bf r}_{2} = {\bf R}_{12} - \frac{1}{2} {\bf 
r}_{12}
\end{equation}
The momenta of the two protons are now given by

\begin{equation}\label{36}
{\bf p}_{1} = M \frac{d{\bf r}_{1}}{dt} = M 
\frac{d{\bf R}_{12}}{dt} + \mu_{12} \frac{d {\bf 
r}_{12}}{dt}
\end{equation}

\begin{equation}\label{37}
{\bf p}_{2} = M \frac{d{\bf r}_{2}}{dt} = M 
\frac{d{\bf R}_{12}}{dt} - \mu_{12} \frac{d {\bf 
r}_{12}}{dt}
\end{equation}
where $\mu = \frac{M}{2}$ is the reduced mass of 
the two-proton system.
Adding the above two equations gives

\begin{equation}\label{38}
{\bf p}_{1} + {\bf p}_{2} = 2M \frac{d{\bf 
R}_{12}}{dt} = {\bf P} = -{\bf q}
\end{equation}
Whereas subtraction leads to the relative 
momentum between the two protons

\begin{equation}\label{39}
{\bf p}_{f} = \mu_{12} \frac{d {\bf r}_{12}}{dt} 
= \frac{1}{2} ({\bf p}_{1} - {\bf p}_{2})
\end{equation}
The unpolarized double differential 
cross-section was written in the form 

\begin{equation}\label{40}
\frac{d^2 \sigma_0}{dW d \Omega_f d \Omega} = 
\frac{p_f}{W} \frac{d^2 \sigma_0}{d^3 p_{f} d 
\Omega} = \frac{1}{4} 
\text{tr}[\mathcal{M}\mathcal{M}^\dag]
\end{equation}
where $d\Omega$ and $d\Omega_f$ respectively 
denote the solid angles associated with ${\bf 
q}$ and ${\bf p}_f$ and $\mathcal{M}$ denotes 
the reaction matrix for
$ pp \to pp\pi^0.$
Recalling the form \eqref{16} for $ pp \to 
p\Delta^+,$ we may express $\mathcal{M}$ since 
the spins $s_1$ and $s_2$ of the two nucleons 
combine to give $s_f$.

\begin{eqnarray}\label{41}
|(l^\prime s_1)s_\Delta s_2 ; s m_\Delta \rangle 
= \sum_{s_f = 0}^1 W(l^\prime s_1 s s_2 
;s_\Delta s_f)[s_\Delta][s_f] \cdot \nonumber \\ 
|l^\prime (s_1 s_2) s_f ; s m_\Delta \rangle 
\hspace{0.8in}
\end{eqnarray}
Since the spins $s_1$ and $s_2$ of the two 
nucleons combine to give $s_f$.
It follows from above that if $s_f = 0,$ then 
$s$ can either be $0$ or $1,$ since $l^\prime = 
1.$
However, if $l^\prime = 1$ then $s$  must be 
equal to 1 only.
To identify the connection between the orbital 
angular momentum quantum numbers $l^\prime$ and 
$l_2$ associated with $ pp \to p\Delta^+,$ and 
the orbital angular momentum quantum numbers $l$ 
and $l_f$ associated with $ pp \to pp\pi^0,$ it 
is convenient to express $(Y_{l^\prime} 
(\hat{\bf q^\prime}) \otimes Y_{l_2} (\hat{\bf 
p_2}))^l_m$ in terms of $(Y_{l} (\hat{\bf q}) 
\otimes Y_{l_f} (\hat{\bf p_f}))^l_m .$
Relative momentum ${\bf q}^\prime$ between the 
pion and the nucleon during the decay of 
$\Delta^+ \to p\pi^0$  can be expressed, 
following equations \eqref{15}, \eqref{38} and 
\eqref{39}, in terms of ${\bf q}$ and ${\bf 
p}_f$ as

\begin{equation}\label{42}
{\bf q^\prime} = \alpha {\bf p}_{f} + \beta {\bf 
q}
\end{equation}
where

\begin{equation}\label{43}
\alpha = -\frac{m}{M+m} ; \beta  = \frac{2M+ 
m}{2(M+m)}
\end{equation}
On the other hand, the momentum ${\bf p}_2$ of 
the outgoing proton in $ pp \to p\Delta^+$ can 
be expressed in terms of ${\bf q}$ and ${\bf 
p}_f$ as

\begin{equation}\label{44}
{\bf p}_{2} = - \left( {\bf p}_{f} + \frac{{\bf 
q}}{2} \right)
\end{equation}
We may also express spherical harmonics $Y_{l m} 
(\hat{\bf q})$ in terms of solid harmonics 
$\mathcal{Y}_{l m} ({\bf q})$
so that $Y_{l^\prime m^\prime} (\hat{\bf 
q}^\prime)$ and $Y_{l_2 m_2} (\hat{\bf p}_2)$ 
may be written in the form

\begin{equation}\label{45}
Y_{l^\prime m^\prime} (\hat{\bf q}^\prime) = 
\frac{1}{(q^\prime)^{l^\prime}} \mathcal{ 
Y}_{l^\prime m^\prime} ({\bf q^\prime}) = 
\frac{1}{(q^\prime)^{l^\prime}}
 \mathcal{ Y}_{l^\prime m^\prime} (\alpha {\bf 
p}_{f} + \beta {\bf q})
\end{equation}
and

\begin{eqnarray}\label{46}
Y_{l_2 m_2} (\hat{\bf p}_2) = 
\frac{1}{(p_2)^{l_2}} \mathcal{Y}_{l_2 m_2} 
({\bf p}_2) \hspace{1cm}\nonumber \\
 = \frac{(-1)^{l_2}}{(p_2)^{l_2}}  \mathcal{ 
Y}_{l_2 m_2} \left( {\bf p}_{f} + \frac{{\bf 
q}}{2} \right)
\end{eqnarray}
Using the formula \cite{33}

\begin{eqnarray}\label{47}
\mathcal{Y}_{\lambda \mu} ({\bf a}+{\bf b}) = 
\sqrt{4\pi} \sum^{\lambda}_{L = 0} 
\frac{1}{\sqrt{2L+1}} \left(
\begin{array}{c}
2\lambda +1 \\
2L 
\end{array}
\right)^{\frac{1}{2}} \nonumber \\ ( \mathcal{ 
Y}_L ({\bf b}) \otimes  \mathcal{ Y}_{\lambda - 
L} ({\bf a}))^\lambda_\mu \hspace{1.5cm}
\end{eqnarray}
where $\left(
\begin{array}{c}
2\lambda +1 \\
2L 
\end{array}
\right) = \frac{(2\lambda+1)!}{2L! (2\lambda +1 
-2L)!}.$

Converting the solid harmonics into spherical 
harmonics, we may write

\begin{eqnarray}\label{48}
Y_{l^\prime m^\prime} (\hat{\bf q}^\prime) = 
\sqrt{4\pi}\sum^{l^\prime}_{L = 0} \frac{1}{[L]} 
\left(
\begin{array}{c}
2l^\prime +1 \\
2L 
\end{array}
\right)^{\frac{1}{2}} \nonumber \\
\frac{(\beta q)^L  (\alpha p_f)^{l^\prime - 
L}}{(q^\prime)^{l^\prime}} (Y_L (\hat{\bf q}) 
\otimes Y_{l^\prime - L} (\hat{\bf 
p}_f))^{l^\prime}_{m^\prime}
\end{eqnarray}
Similarly

\begin{eqnarray}\label{49}
Y_{l_2 m_2} (\hat{\bf p}_2) = \sqrt{4\pi} 
(-1)^{l_2} \sum^{l_2}_{L^\prime = 0} 
\frac{1}{[L^\prime]} \left(
\begin{array}{c}
2l_2 +1 \\
2L^\prime 
\end{array}
\right)^{\frac{1}{2}}
\nonumber \\
\frac{( q)^{L^\prime}  (p_f)^{l_2 - 
L^\prime}}{(2)^{L^\prime} p_2^{l_2}} 
(Y_{L^\prime} (\hat{\bf q}) \otimes Y_{l_2 - 
L^\prime} (\hat{\bf p}_f))^{l_2}_{m_2} \quad
\end{eqnarray}
So that

\begin{eqnarray}\label{50}
(Y_{l^\prime} (\hat{\bf q}^\prime) \otimes 
Y_{l_2} (\hat{\bf p}_2))^{L_f}_{M_f} = \sum_{L, 
L^\prime, l, l_f} \mathcal{G} \hspace{1.25in} 
\nonumber \\
\left((Y_L (\hat{\bf q}) \otimes Y_{l^\prime - 
L} (\hat{\bf p}_f))^{l^\prime} \otimes 
(Y_{L^\prime} (\hat{\bf q}) \otimes Y_{l_2 - 
L^\prime} (\hat{\bf p}_f))^{l_2}
\right)^{L_f}_{M_f} \quad 
\end{eqnarray}
where $\mathcal{G}$ is a factor free from 
angular dependance.

Re-coupling the tensor products involving 
spherical harmonics to bring the spherical 
harmonics with arguments $\hat{\bf q}^\prime s$ 
and $\hat{\bf p}_f^\prime s$
together and combining the two spherical 
harmonics with same arguments leads to

\begin{eqnarray}\label{51}
(Y_{l^\prime} (\hat{\bf q}^\prime) \otimes 
Y_{l_2} (\hat{\bf p}_2))^{L_f}_{M_f} = 
\sum_\zeta F \, C(L L^\prime l;000)\nonumber \\ 
C(l^\prime - L,l_2 - L^\prime, l_f;000) \left( 
Y_l (\hat{\bf q}) \otimes Y_{l_f} (\hat{\bf 
p}_f) \right)^{L_f}_{M_f} \nonumber \\
 \left\{
\begin{array}{ccc}
L & L^\prime  & l \\
l^\prime-L & l_2 - L^\prime & l_f \\
l^{\prime} & l_2 & L_f
\end{array}
\right\}
\end{eqnarray}
where the summation is carried out over the set 
of indices denoted by

\begin{equation}\label{52}
\zeta = \{ L, L^\prime,  l, l_f \}
\end{equation}
and the factor $F$ is given by

\begin{eqnarray}\label{53}
F = \frac{(-1)^{l_2}}{(2)^{L^\prime} p_2^{l_2} 
(q^\prime)^{l^\prime}}\: \beta^L 
\alpha^{l^\prime - L} q^{L+L^\prime}
p_f^{l_2 - L^\prime + l^\prime - L} \nonumber \\
([l^\prime][l_2][l^\prime - L][l_2 - L^\prime])
\left( \begin{array}{c}
2l^\prime +1 \\
2L 
\end{array}
\right)^{\frac{1}{2}} \left(
\begin{array}{c}
2l_2 +1 \\
2L^\prime
\end{array}
\right)^{\frac{1}{2}}
\end{eqnarray}
Since the decay of $\Delta$ into $p\pi^0$ is a 
$p-$wave, $l^\prime = 1.$
$l_2$ can take values $0$ or $1.$
In particular if we take $l_2 = 1$ the 
\eqref{51} reduces to the form

\begin{eqnarray}\label{54}
(Y_1(\hat{\bf q}^\prime) \otimes Y_1 (\hat{\bf 
p}_2))^{L_f}_{M_f} = \frac{-1}{q^\prime p_2}[ 
\frac{3}{[L_f]} C(11L_f; 000) \nonumber \\
\left\{ \alpha p_f^2 ( Y_0 (\hat{\bf q}) \otimes 
Y_{L_f} (\hat{\bf p}_f))^{L_f}_{M_f} + 
\frac{\beta q^2}{2} ( Y_{L_f} (\hat{\bf q} 
\otimes Y_0(\hat{\bf p}_f))^{L_f}_{M_f} \right\}  
\nonumber \\
+ \left( \frac{(-1)^{L_f}\alpha}{2}+  \beta 
\right) 
p_f q ( Y_1 (\hat{\bf q}) \otimes Y_{1} 
(\hat{\bf p}_f))^{L_f}_{M_f}] \hspace{1cm}
\end{eqnarray}
with $L_f = 0$ 

\begin{eqnarray}\label{55}
(Y_1(\hat{\bf q}^\prime) \otimes Y_1 (\hat{\bf 
p}_2))^{0}_{0} =  \frac{-1}{q^\prime p_2}   
[\sqrt{3}\alpha p_f^2  + \frac{\sqrt{3}\beta 
q^2}{2}]\nonumber \\ ( Y_0 (\hat{\bf q}) \otimes 
Y_{0} (\hat{\bf p}_f))^{0}_{0}
 + \left( \frac{\alpha}{2}+ \beta \right) p_f q 
( Y_1 (\hat{\bf q}) \otimes Y_{1} (\hat{\bf 
p}_f))^{0}_{0} \hspace{1cm}
\end{eqnarray}
with $L_f = 1$ 

\begin{equation}\label{56}
(Y_1(\hat{\bf q}^\prime) \otimes Y_1 (\hat{\bf 
p}_2))^{1}_{M_f} =  \frac{1}{q^\prime p_2}\left( 
\frac{\alpha}{2} - \beta \right) p_f q
( Y_1 (\hat{\bf q}) \otimes Y_{1} (\hat{\bf 
p}_f))^{1}_{M_f}
\end{equation}
with $L_f = 2$

\begin{eqnarray}\label{57}
(Y_1(\hat{\bf q}^\prime) \otimes Y_1 (\hat{\bf 
p}_2))^{2}_{M_f} =  \frac{-1}{q^\prime p_2}[ 
\frac{3\alpha p_f^2}{\sqrt{5}} ( Y_0 (\hat{\bf 
q}) \otimes Y_{2}
(\hat{\bf p}_f))^{2}_{M_f} \hspace{1cm}\nonumber 
\\
+ \frac{3\beta q^2}{2\sqrt{5}} ( Y_{2} (\hat{\bf 
q} \otimes Y_0 (\hat{\bf p}_f))^{2}_{M_f} 
\hspace{2cm} \nonumber \\
 + \left( \frac{\alpha}{2} + \beta \right) p_f q 
( Y_1 (\hat{\bf q}) \otimes Y_{1} (\hat{\bf 
p}_f))^{2}_{M_f} ] \hspace{1cm}
\end{eqnarray}

\section{RESULTS AND CONCLUSIONS}

\hspace{0.5cm} The two Clebsch-Gordan 
coefficients and $9j$ symbol in \eqref{51} allow 
us to choose appropriate $l$ and $l_f$ 
corresponding to given $l_2.$
With $l_2 = 0,1$ we have a set of nine partial 
wave amplitudes $F_i$ with $i = 1,2, \ldots, 9$ 
for $pp \to p\Delta^+$ listed in
Table$-II$ which can contribute to those partial 
wave amplitudes $f_j$ with $j = 1,2, \ldots, 16$ 
for the reaction $pp \to
pp\pi^0$ listed in Table$-III$ which is 
consistent with \eqref{51}.

If $l_2 = 0$, then only possible solution 
corresponds to $l = 0$ and $l_f = 1.$
However, if $l_2 = 1$, then allowed values of 
$l$ and $l_f$ are\\
$i)$ $ l = 0$ and $l_f = 0;$\\
$ii)$ $ l = 1$ and $l_f = 1$\\
$iii) $ $l = 0$ and $l_f = 2$\\
$iv)$ $l = 2$ and $l_f = 0.$

Taking into consideration the spin, orbital 
angular momentum and the total angular momentum 
conservation, the connection between
the two sets of partial wave amplitudes $F_i$ 
with $i = 1,2, \ldots, 9$ and $f_j$ with $j = 
1,2, \ldots, 16$ can be established.
The Table-IV given below shows the connection 
between the two sets of partial wave amplitudes.

\begin{table}[h]
\caption{Connection between $\Delta$ production 
and $pp\pi^0$ amplitudes}
\begin{tabular}{|cccccccc|}\hline
$l_2$ & $s$ &  $j$ &  $F_i$ & $l$ & $l_f$ & 
$s_f$ & $f_j$ \\ \hline
0 & 2 & 2 & $F_1$ & 0 & 1 & 1 & $f_3$ \\ \hline
1 & 1 & 0 & $F_2$ & 0 & 0 & 0 & $f_1$ \\
 &  &  & & 1 & 1 & 1 & $ f_4$ \\ 
 &  & 1 & $F_3$ & 1 & 1 & 1 & $ f_9 , f_{10} , 
f_{11}$ \\
 &  & 2 & $F_4 , F_5$ & 1 & 1 & 1 & $ f_5 , f_6 
, f_7 , f_8 $ \\
 &  &  & & 0 & 2 & 0 & $ f_{13}, f_{14}$ \\
 &  &  & & 2 & 0 & 0 & $ f_{15}, f_{16}$ \\ 
 & 2 & 1 & $F_6$ & 1 & 1 & 1 & $f_9 , f_{10}, 
f_{11}$ \\
&  & 2 & $F_7 , F_8$ & 1 & 1 & 1 & $ f_5 , f_6 , 
f_7 , f_8 $ \\ \hline
\end{tabular}
\end{table}
It is seen from Table-IV that\\
$i)$ $Ss$ amplitude derive contribution from 
$F_2$ only.\\
$ii)$ $Ps$ amplitudes derive contributions from 
$F_1, F_2, F_4, F_5$ and $ F_7.$\\
$iii)$ $Pp$ amplitudes will derives 
contributions from $F_2, F_8.$\\
$iv)$$Sd$ and $Ds$ amplitudes derive 
contributions from $F_4$ and $F_5$.

\section{Acknowledgement}

\hspace{0.5cm} One of the authors, Venkataraya 
thanks Dr. S. Mahadevan, Professor and Dean, 
Department of Sciences, Amrita Vishwa 
Vidyapetham, Coimbatore for his encouragement to 
research work. The authors thank Dr. Shilpashree 
S P and Venkataramana Shastri for  helpful 
discussions.


\begin{thebibliography}{99}

\bibitem{1} Clark, Roberts and Wilson, Phys. 
Rev. $\bf83$ 649 (1951); Durbin, Loar and 
Steinberger, Phys. Rev. $\bf84$, 581 (1951);
J.Marshall; L.Marshall, V.A.Nedzel and 
S.D.Warshaw, Phys.Rev.$\bf88$, 275 (1952); 
R.H.Hildebrand, Phys. Rev. $\bf89$, 1090 (1953);
J.W.Mather and E.A.Martinelli, Phys. Rev. 
$\bf92$, 780 (1953).

\bibitem{2} H.Yukawa, Proc.Phys.Math.Soc.Japan 
$\bf17$, 48 (1935)

\bibitem{3} C.M.G Lattes, H.Muirhead, C.F.Powell 
and G.P.S.Gcchialini, Nature $\bf159$, 694 
(1947);
C.M.G.Lattes, G.P.S.Gcchialini and C.F.Powell, 
Nature $\bf160$, 453 (1947); A.G.Carlson, 
J.E.Hooper and D.T.King, Phil.Mag.$\bf41$, 701 
(1950);
R.Bjorkland, W.E.Crandell, B.J.Moyer and 
H.F.York, Phys. Rev. $\bf77$ 2131 (1950).

\bibitem{4} K.Brueckner, Phys.Rev.$\bf82$, 598 
(1951); K.M.Watson and K.Brueckner, Phys. Rev. 
$\bf83$, 1 (1951);
G.F.Chew, M.L.Goldberger, Steinberger and 
C.N.Yang, Phys. Rev. $\bf84$, 581 (1951).

\bibitem{5} K.M.Watson and C.Richman, Phys. Rev. 
$\bf83$, 1256 (1951); R.E.Marshak and 
A.M.L.Messiah, Nuov Lim $\bf11$, 337 (1954).

\bibitem{6} A.H.Rosenfeld, Phys. Rev. $\bf96$, 
130, 139 (1954); M.Gellmann and K.M.Watson, Ann. 
Rev. Nucl. Sc. $\bf4$, 219 (1954).

\bibitem{7} R.E.Pollock, Ann. Rev. Nucl. Part. 
Sc. $\bf41$, 357 (1991).

\bibitem{8} H.O.Meyer {\it et al.}, Phys. Rev. 
Lett $\bf65$, 2846 (1990);
Nucl. Phys. $\bf A539$, 633 (1992); R.Beck {\it 
et al.}, Phys. Rev. Lett $\bf65$, 1841 (1990);
S.Stanislanes {\it et al.}, Phys. Rev. $C41$, 
1913 (1990); D.A.Hutcheo {\it et al.}, Nucl. 
Phys. 618 (1991);
A.Bondar {\it et al.}, Phys. Lett B, $\bf356$, 8 
(1995).

\bibitem{9} W.W.Dachnick, Phys. Rev. Lett 
$\bf74$, 2913 (1993).

\bibitem{10} E.Korkmaz {\it et al.}, Nucl. Phys 
$\bf A535$, 636 (1991);
D.A.Hutchson {\it et al.}, Nucl. Phys. $\bf 
A535$, 618 (1991);
M.Drochner {\it et al.}, Phys. Rev. Lett 
$\bf77$, 454 (1996);
P.Heimberg {\it et al.}, Phys. Rev. Lett 
$\bf77$, 1012 (1996).


\bibitem{11} D.S.Koltun and A.Reitan, Phys. Rev. 
$\bf141$, 1413 (1966);
Nucl. Phys. $\bf B4$, 629 (1968);
S.I.Adler and R.F.Dashen, Current algebras and 
their applications to particle physics 
(Benjamin, New York 1968);
M.E.Schillaci, R.R.Silbar and J.E.Young, Phys. 
Rev. Lett $\bf21$, 711 (1968); Phys. Rev. 
$\bf179$, 1539 (1969);
M.E.Schillaci and R.R.Silbar, Phys. Rev. 
$\bf185$, 1835 (1969); D.O.Riska {\it et al.}, 
Phys. Lett. $\bf61B$, 41 (1976);
Ch.Weddigen, Nucl. Phys. $\bf A312$, 330 (1978); 
J.Gasser and H.Leutiwgler, Phys. Rep. $\bf87$, 
77 (1982);
T.S.H.Lee and A.Matsuyama, Phys Rev $\bf C36$, 
1459 (1987); H.Garilazo and T.Mizutani, $\pi$NN 
systems (World Scientific, Singapore, (1990).

\bibitem{12} G.A. Miller and P.U.Sauer, Phys Rev 
$\bf C44$, R1725 (1991); J.A.Niskanen, Phys.Lett 
$\bf B289$, 227 (1992);
T.S.H.Lee and D.O.Riska, Phys Rev Lett $\bf 70$, 
2237 (1993); C.Horowitz, H.O.Meyer and 
D.Greiger, Phys Rev $C49$, 1337 (1993);
C.J.Horowitz, Phys Rev $\bf C48$, 2920 (1993).

\bibitem{13}  
H.O.Meyer,\,Ann.\,Rev.\,Nucl.\,Part.\,Sc.\,$\bf{
47}$, 235 (1997); F.Rathmann {\it et al.}, 
Phys.\,Rev.\,$\bf{C58}$, 658 (1997).

\bibitem{14}  H.O.Meyer\, {\it et al.}, Phys. 
Rev. Lett. $\bf {81}$, 3096 (1998); Phys. Rev. 
Lett. $\bf {83}$, 5439 (1999).

\bibitem{15} J.Haidenbauer, Ch.Hanart and 
J.Speth, Aefce Phys.Pol. $\bf {B27}$, 2893 
(1996);Ch.Hanhart, Ph.D Thesis, Bonn University 
(1997);
Ch.Hanhart, J.Haidenbauer, O.Krehl and 
J.Speth,Phys.Lett. $\bf {B444}$, 25 (1998).

\bibitem{16} S.M.Bilenky and R.M.Ryndin, Phys. 
Lett. $\bf 6$, 217 (1963).

\bibitem{17} J.Zlomanczak {\it et al.}, Phys. 
Lett. $\bf B436$, 251 (1998);
J. A. Niskanen, Phys. Lett. $\bf {B289}$, 227 
(1992): Y. Maeda et al,  N News Letter $\bf 
{13}$, 326 (1997);
B. Bilger et al., Nucl. Phys.$\bf {A 693}$, 633 
(2001).

\bibitem{18} V.Bernard, N.Kaiser and 
U-G.Mei$\beta$ner, Int. J. Mod. Phys. $\bf{E4}$, 
193 (1995);
H.Machner and J.Haidenbauer J. Phys. G: Nucl. 
Part. Phys. $\bf{25}$ R231 (1999).
P. Moskal, M. Wolke, A. Khoukaz, and W. Oelert 
Prog. Part. Nucl. Phys. $\bf{49}$, 1 (2002);
G. Fäldt, T. Johannson and C. Wilkin, Phys. 
Scripta $\bf{T99}$,146 (2002).
C. Hanhart, Phys. Rep. $\bf{397}$, 155 (2004).

\bibitem{19} G.Ramachandran, P.N.Deepak and 
M.S.Vidya Phys. Rev. $\bf {C62}$, 011001(R) 
(2000).

\bibitem{20} G.Ramachandran and P.N.Deepak, 
Phys. Rev. $\bf {C62}$, 051001(R) (2001).

\bibitem{21} H.O.Meyer {\it et al.}, Phys. Rev. 
$\bf {C63}$, 064002 (2001).

\bibitem{22} B.v.Przewoski et al., \prc 
$\bf{61}$, 064604 (2000);W.W.Daehnick et al., 
\prc $\bf{65}$, 024003(2002).

\bibitem{23} G.Ramachandran and M.S.Vidya, \prc 
$\bf{56}$, R12 (1997).

\bibitem{24} P.N.Deepak and G.Ramachandran, 
Phys. Rev. $\bf {C65}$, 027601 (2002);
P.N.Deepak, G.Ramachandran and C.Hanhart, 
Matter. Mater $\bf 21$, 138 (2004);
P.N.Deepak, C.Hanhart, G.Ramachandran and 
M.S.Vidya, Int. J. Mod. Phys. $\bf A20$, 599 
(2005).

\bibitem{25} P.N.Deepak, J.Haidenbauer and 
C.Hanhart Phys. Rev. $\bf {C72}$, 024004(2005).

\bibitem{26} G.Ramachandran, G.Padmanabha, and 
Sujith Thomas \prc $\bf{81}$, 067601 (2010).

\bibitem{27} G.Glass {\it et al.}, Phys. Rev. 
$\bf {D15}$, 36 (1977); Phys. Lett. $\bf 
{B129}$, 27 (1983)\\
F.Skimizu {\it et al.}, Nucl. Phys $\bf {A386}$, 
571 (1982);  Nucl. Phys $\bf {A389}$, 445 
(1982)\\
A.D.Hancock {\it et al.}, Phys. Rev. $\bf 
{C27}$, 2742 (1983)\\
T.S.Bhatia {\it et al.}, Phys. Rev. $\bf {C28}$, 
2671 (1983)\\
B.K.Jain, Phys. Rev. Lett. $\bf 50$, 815 
(1983)\\
W.W.Dachnick {\it et al.}, Phys. Rev. Lett. $\bf 
74$, 2913 (1995)\\
N.G.Kelkar and B.K.Jain Int. J. Mod. Phys. 
$\bf{E4}$, 181 (1995)
J.G.Hardie {\it et al.}, Phys. Rev $\bf {C56}$, 
20 (1997).

\bibitem{28} R. R. Silbar, R. J. Lombard, and W. 
M. Kloet, Nucl. Phys. $\bf{A381}$,381 (1982).

\bibitem{29} L. Ray, Phys. Rev. C $\bf{49}$, 
2109 (1994).

\bibitem{30} J.P.Auger and C.Lazard Phys.Rev. C 
$\bf{52}$, 513(1995).

\bibitem{31} Y.Maeda {\it et al.}, Phys. Rev 
$\bf {C77}$, 044004 (2008).

\bibitem{32} A.Budzanowski {\it et al.}, Phys. 
Rev $\bf {C79}$, 061001 (2009)\\
V.Baru, E.Epelbaum, J.Haidenbauer, C.Hanhart, 
A.E.Kudryatsev, V.Lensky, and U-G.Mei$\beta$ner 
\prc $\bf 80$, 044003 (2009).

\bibitem{33} M. E. Rose, J. Math. Phys., {\bf 
37}, 215  (1958)

\end{thebibliography}
\end{document}